\documentclass[sigconf, nonacm]{acmart}
\usepackage{subcaption}

\AtBeginDocument{%
  }

\copyrightyear{2025}
\acmYear{2025}
\setcopyright{rightsretained}
\acmConference[CHI EA '25]{Extended Abstracts of the CHI Conference on Human Factors in Computing Systems}{April 26-May 1, 2025}{Yokohama, Japan}
\acmBooktitle{Extended Abstracts of the CHI Conference on Human Factors in Computing Systems (CHI EA '25), April 26-May 1, 2025, Yokohama, Japan}\acmDOI{10.1145/3706599.3719989}
\acmISBN{979-8-4007-1395-8/2025/04}

\begin{document}

\title{Social Media Journeys – Mapping Platform Migration}

\author{Artur Solomonik}
\orcid{0009-0000-8349-148X}
\affiliation{%
  \institution{Center for Advanced Internet Studies (CAIS) gGmbH}
  \city{Bochum}
  \country{Germany}
}
\email{artur.solomonik@cais-research.de}
\author{Hendrik Heuer}
\affiliation{%
  \institution{Center for Advanced Internet Studies (CAIS) gGmbH}
  \city{Bochum}
  \country{Germany}
}
\affiliation{%
  \institution{University of Wuppertal}
  \city{Wuppertal}
  \country{Germany}
}
\email{hendrik.heuer@cais-research.de}

\renewcommand{\shortauthors}{Solomonik and Heuer}

\begin{abstract}
As people engage with the social media landscape, popular platforms rise and fall. As current research uncovers the experiences people have on various platforms, rarely do we engage with the sociotechnical migration processes when joining and leaving them. In this paper, we asked 32 visitors of a science communication festival to draw out artifacts that we call Social Media Journey Maps about the social media platforms they frequented, and why. By combining qualitative content analysis with a graph representation of Social Media Journeys, we present how social media migration processes are motivated by the interplay of environmental and platform factors. We find that peer-driven popularity, the timing of feature adoption, and personal perceptions of migration causes—such as security—shape individuals' reasoning for migrating between social media platforms. With this work, we aim to pave the way for future social media platforms that foster meaningful and enriching online experiences for users.
\end{abstract}



\maketitle

\section{Introduction}
After a period of relative stability of the 2010's social media landscape, digital migrations have become increasingly common, as evidenced by the exodus from Twitter (now X)~\cite{HuangMastodon2022}. While popular platforms have progressively replaced early platforms from 30 years ago, their lasting influence on what people desire from social media remains largely unexplored. Furthermore, the overarching factors that drive users to choose one platform over another remain understudied within the broader context.

Social media audiences are highly distributed across various platforms, with less than one in twenty platform users (4.5\%)  being unique to that platform~\cite{meltwaterSocialMediaStatistics2024}. This widespread distribution highlights the transient and exploratory nature of online engagement. As new platforms emerge, evolve, and close, users continually seek online spaces that align with their shifting needs~\cite{haimsonTransitionMachinery2018, haimsonTumblrWasTrans2021}, enabling discussions about the physical world context~\cite{semaanNavigatingAudience2015}.

Prominent design paradigms often fail to accommodate non-normative social media usage~\cite{kenderBanalAutistic2023}, even though unique digital spaces can offer features supporting atypical modes of sociality ~\cite{ringlandMinecraft2016}. In this paper, we operationalize the term digital migration as the act of moving from one platform to another for any reason. The sequence of migrations of a person we consider their Social Media Journey. Understanding the motivations driving individuals to adopt and migrate between social media platforms—and the paths they take—provides valuable insights into what users prioritize in the digital landscape.

This paper contributes insights on how motivations for social media migration have evolved over time, with a particular focus on the interplay between platform affordances, design features, and social norms. By investigating the factors that users retrospectively identify as reasons for their migration, we foreground the dynamic relationship between platform design principles, social contexts, and the perceptions shaping the social media landscape. We address the following two research questions:

\begin{enumerate}
    \item [RQ1] What platform affordances, design features, and social norms do people retrospectively self-report as reasons to migrate between social media platforms?
    \item [RQ2] How have the reasons for social media migration changed over time?
\end{enumerate}

To address these questions, we invited science communication festival visitors to map and elaborate on their Social Media Journeys, contextualizing migration factors within space and time. Drawing on prior work by HCI scholars, we expand the set of factors contributing to platform migration, incorporating the context of different internet eras. This approach highlights unique and meaningful migration processes that might otherwise go unnoticed in broader discussions. With this paper, we contribute 1) a complication of prior work on social media migration from a human-centered perspective, introduce 2) Social Media Journey Maps as a novel method to collect qualitative data, and 3) highlight the influence of niche or popular social media on navigating the social media landscape.

\section{Background}

Our work draws on recent research on social media spaces and how they relate to digital migration. We situate our work among CSCW and CHI scholars~\cite{schlesingerSituatedAnonymityImpacts2017, haimsonTransitionMachinery2018, fieslerMoving2020, zhangFormFrom2024}, who provide an insightful foundation for working with social media platform affordances and social media graphs. 

\subsection{Social Media Taxonomies}

We adopt the Form-From framework described by Zhang et al. to characterize platforms based on the shape that content takes while considering its source~\cite{zhangFormFrom2024}. With this suitable abstraction of social media infrastructure, we can confront explicit platforms with the social contexts they intersect with.

Taking into account the role of a platform's members, Ouridi at al. propose a classification of social media platforms based on their users, content format, and function~\cite{ouirdiSocialMediaConceptualization2014}. In a step further, Santos highlights the role of popular social media as significant political and economic actors~\cite{santosSocalledUGCUpdated2022}. While these definitions provide a foundational understanding of social media in a societal context, DeVito complicates these taxonomies by introducing users' folk theories about platform algorithms, and considering how platforms have evolved over time~\cite{devitoAdapt2021}. As such, HCI research offers more nuanced ways to engage with the definition of social media platforms, allowing for a meaningful confrontation with emerging technologies.

Direct analyses of the realities on respective social media platforms offer a chance to specify sociotechnical concepts. Schlesinger et al. categorize the pseudonymous social media application Yik Yak by its anonymity, ephemerality, and hyper-locality~\cite{schlesingerSituatedAnonymityImpacts2017}. Similarly, Haimson et al. investigate how platforms like Tumblr navigate the challenges of supporting individuals undergoing gender transition, uncovering the intricate interplay between platform design and user experiences~\cite{haimsonTransitionMachinery2018}. While Schlesinger et al. focus on the social complexities emerging from platform design, Haimson et al. explore how these complexities intersect with the liminal experiences of users. With this duality of platform features and fluid identity in mind, our work aims to capture how people confront such platforms when navigating the social media landscape.

Building on the Form-From model in relation to how Schlesinger et al. and Haimson et al. characterize social media, we situate the platforms mentioned by our participants within their Social Media Journeys. By doing so, we explore platform design as a contributing factor to migration and its role in the broader sociotechnical context. Our research aims to examine such intersections to understand what social media platforms have afforded different users throughout the history of social media. 

\subsection{Reasons for choosing a Social Media Platform}

To understand what contributes to choosing social spaces to engage with, examining the realities of existing platforms is crucial. Analyzing the user bases of platforms prior to the advent of Facebook, Hargittai underpins how gender, race, ethnicity and educational background contribute to which platforms are adopted by users~\cite{hargittaiWhoseSpaceDifferences2007}. DeVito et al. introduce the concept of personal social media ecosystems that situate platforms alongside audiences, affordances, and norms for people to navigate. We consider affordances in our context to be shaped by the interactions between artifacts, users, and social contexts, as well as their political and negotiated dimensions~\cite{davisAffordances2020}. We aim to understand how such aspects may have influenced a local social media landscape and how it shaped the people it once attracted.

Our work heavily draws from the insights of scholars of Queer HCI who offer valuable critical insight into engaging with social media research as a whole\cite{haimsonTransitionMachinery2018, deVitoTooGay2018, fieslerMoving2020, kenderBanalAutistic2023}. As a distinct case for non-normative perspectives on social media, Haimson et al. describe how trans people felt exceptionally drawn to the platform Tumblr. That special case served as a suitable space for trans people to experiment with their gender expression outside of their social environment, hence aligning with the very specific needs trans people may have when engaging with communities online~\cite{haimsonTumblrWasTrans2021}.

Accordingly, we aim to account for factors of locality and residence, and how they coincide with social media platforms. By exploring especially small platforms and communities, we try to capture the nuance when talking about the time and space of a platform. As first of its kind, our work features the unique social media landscape history of Germany to broaden our understanding on why people gravitate towards certain social media platforms.

\subsection{Digital Migration}

As a first step into finding patterns within acts of migration, Kumar et al. offer an intuition by monitoring the user attention of users~\cite{kumarUserMigrationPatterns2011}. With their quantification of users migrating away from popular platforms in mind, we aim to expand the variety of influences involved in processes of platform migration. As such, we follow a human-centered approach as opposed to monitoring selected social media platforms.

Fiesler and Dym highlight fandom communities and how they are influenced by user migration~\citet{fieslerMoving2020}. They stress that even beyond issues of policy and value, migration between platforms is complex at its core. Even when emerging spaces seem more appealing, the adoption by friends and family is what can strongly influence a platform's status as a new home. This unique perspective on fandom communities offers great insight into people of similar migration goals, and gives us the opportunity to pursue migration strategies from a generalizing angle.

Building on the complexities of locality and temporality when situating social media platforms, our data revealed unique perspectives on how our local social media ecosystem has changed over time. 

\section{Methods}

We conducted a qualitative study at a science communication event in Germany. We chose this particular venue for the opportunity to invite participants of a broad age range, its local experiences and interest in contributing to social media research. We asked people about the platforms each person frequented and why they chose to leave or join them with the help of their Social Media Journey Map. Participants at the event did not receive any material compensation, but were invited to take part in the numerous games, workshops and presentations including an open buffet.

After obtaining informed consent, participants were asked to reflect on their initial interactions with social media platforms. We provided a large blank sheet of paper and pens on a large desk at the venue. We prepared sixteen types of stickers of the logos of platforms included in the Digital News Report 2021~\cite{newmanReuters2021}. We also provided blank stickers so that participants could add other platforms.

\begin{figure*}
    \centering
    \includegraphics[width=\textwidth]{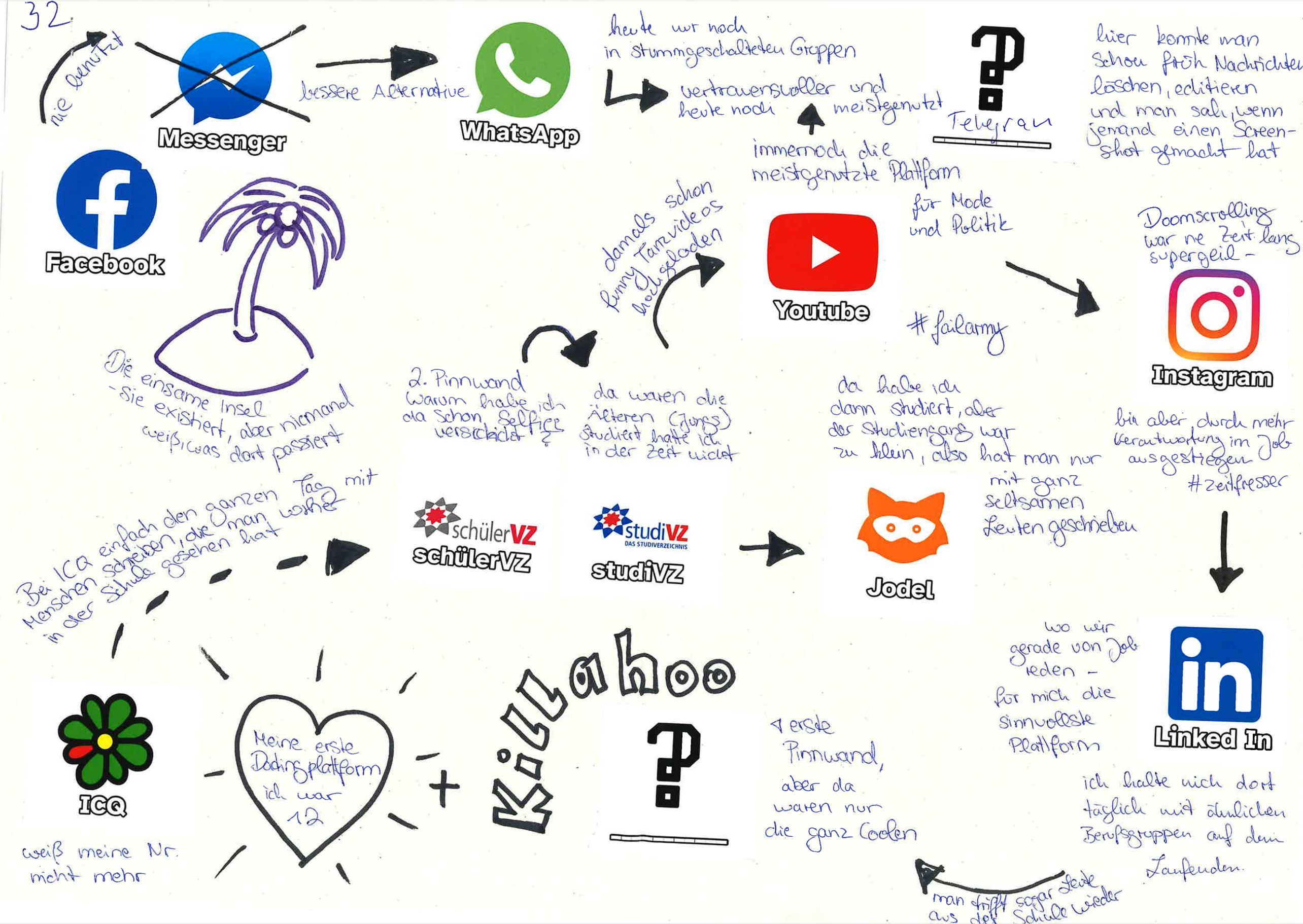}
    \Description{Collage of social media platform logos with handwritten notes and arrows connecting them.}
    \caption{Social Media Journey Map by P9}
    \label{fig:example-drawing}
\end{figure*}

Participants were tasked to recall previously frequented platforms and services. They drew connections to the platforms remembered, and annotated their reason to do so, creating a Social Media Journey Map similar to Figure ~\ref{fig:example-drawing}. These graphs were inspired by qualitative visualization methods, particularly concept mapping~\cite{novakConceptMappingOrigins2006}, and extended by timeline mapping~\cite{kolarTimelineMapping2015} to account for notions of time passing throughout the journey. This variation of concept mapping was chosen for its flexibility, rapid information capture, and focus on relationships among items~\cite{epplerComparisonConceptMap2006}. 

We analyzed the graph drawings, and causes of migration in particular, using qualitative content analysis~\cite{mayringQualitativeContentAnalysis20045}, a method similar to thematic analysis~\cite{braunThematicAnalysis2006}. Based on axial coding principles~\cite{corbinGroundedTheory2014}, the first author observed all graphs multiple times. In weekly meetings with the second author, codes were discussed and clustered to establish categories and sub-categories. After iteratively coding and discussing the data, we obtained our final set of categories. We operationalize the reported platforms by situating them within the ~\cite{zhangFormFrom2024} framework to capture their general mechanisms throughout the discussion.

\subsection{Participants}

The sample consisted of $n=32$ participants (20 women, 6 men, 6 undisclosed), with the majority falling within the age range of 26 to 40 years. The participants' mean age was 32.8 years (SD = 7.9). The majority had Germany’s secondary school leaving certificate, equivalent to A-levels or the International Baccalaureate, granting university eligibility ($37.5\%$, $n=12$), Bachelor's Degree ($9.4\%$, $n=3$) and Master's Degree ($28.1\%$, $n=9$).

\section{Results}

Based on the emerging categories and themes identified within the social media graphs, we present the key factors participants reported as influencing their movement between platforms (RQ1) and how these factors evolved throughout their social media journeys (RQ2).

\subsection{Environmental Motivators}

\subsubsection{Popularity}

Popularity emerged as one of the most prominent factors influencing platform migration, particularly in transitions from early platforms such as MSN, MySpace, and ICQ to Facebook. P6 described how “[Facebook] was used more by the general public” when they decided to move from the now-defunct platform SchülerVZ~\cite{schülervz2013}. As a popular platform with the participants, SchülerVZ was a German community for school students offering profiles, friend lists, messaging, and group forums~\cite{bagerSVZ2008}, until it was terminated in 2013. Conversely, P1 noted that “no one is on Facebook anymore” when explaining their move to Instagram and preference for Reddit, commenting that “nothing is happening [on Facebook]”. Similarly, P9 describes Facebook as ``the lonely island - it exists, but no one knows what is going on there'', while the rest of their journey is described in detail (Figure~\ref{fig:example-drawing}). Facebook was consistently the only platform participants explicitly described as unfashionable, or “out”~(P7, P26). Participants noted leaving early social media platforms~(P5, P15, P35) and emerging platforms like Mastodon and Bluesky~(P1, P4) as they weren't populated enough.

When discussing popularity, participants distinguished between two dimensions: a platform’s general popularity~(P6, P23, P26) and its popularity within their immediate social circles, such as friends or family~(P2, P8, P10). For example, joining Facebook was associated with “becoming more mature” and following their friend group~(P26), while moving to another country required adopting the platform to connect to new peers~(P27).

\subsubsection{Communication Goals}

A common motivation for platform migrations, as reflected in the data, is the desire to communicate with specific groups of people. To stay in touch with international contacts, most participants preferred Facebook~(P3, P4, P6). P2 observed that Snapchat was where “all international teenagers [communicated]”, alluding to their involvement with the platform during a younger age. Platforms like Messenger~(P10) and WhatsApp~(P19) were commonly cited as suitable for family communication, while interactions with friends were distributed across a variety of platforms. For instance, participants reported chatting with school friends on ICQ~(P9), “collecting flames [on Snapchat]” as one would reciprocate photo messages on a daily basis~(P10), or sharing more personal glimpses of their lives on BeReal~(P10).
 
\subsubsection{Work Context}

Some participants also chose to annotate work-related platforms~(P12, P15, P28), noting that they often could not directly connect these to other platforms. LinkedIn particularly stood out as “the most meaningful platform [for work]”~(P32). Additionally, Twitter, Facebook, and Xing emerged as the most popular work platforms, with varying degrees of exclusivity. In particular, some participants used these platforms solely for work purposes, while others combined professional and personal use. Additionally, participants with specific interests often distinguished the platforms they used based on the audience. For example, some turned to particular platforms for discussing activist work~(P23), disseminating work-related content~(P29), or connecting with university friends~(P19). These choices reflect a deliberate distinction of communication by audience and purpose.

\subsection{Platform Motivators}

\subsubsection{Functionality and Features}

Security concerns were a key factor in migrations to messaging applications, particularly when moving from WhatsApp to platforms like Telegram~(P1, P9, P27), Signal~(P3, P8, P15), and Threema~(P14, P15). P1 joined Telegram “because of data security” but later switched to Signal, regarding it as “[more secure] than Telegram.”

Migrations to Instagram were often driven by its introduction of new features, in addition to its visual appeal~(P3, P11). P24 shared that they moved from Instagram “when Instagram introduced Stories, because no one posted there anymore,” highlighting a feature that seemed more suited to Instagram than to its original platform Snapchat~\cite{hamburgerSnapchat2013}.

Participants also reported about features from early social media platforms, such as the group feature on SchülerVZ~(P2), the first pinboards on the small local chat platform serving as a guestbook Kilahu~(P9), and the video chats on Google+~(P12). The group feature of SchülerVZ in particular was a unique concept where one could showcase their affiliation with certain groups similar to fanlistings~\cite{Fanlistings2000} and website banners. However, they were widely adopted as humorous phrases and puns that would decorate each member's user profiles.

\subsubsection{Published Content}

When discussing a platform's content, participants distinguished between engaging with platforms offering a large quantity of content and those focused on specific topical matters. Platforms like Instagram~(P1), YouTube~(P23), and Twitter~(P25) were mentioned for their vast and entertaining user-generated content. Conversely, platforms such as Pinterest~(P2), Reddit~(P23), and earlier platforms or forums~(P16) were primarily valued for their topical focus and community-driven nature.

The term ``doomscrolling''~(P8, P23) and its negative connotations~(P16) frequently arose in discussions about migrations to Instagram and resurfaced with the emergence of TikTok. P16 described having ``no words'' to convey their feelings about TikTok, while P30 even admitted to being ``a little afraid of [it]''. Since doomscrolling is attributed to the infinite feed feature as a dark pattern~\cite{roffarelloDark2022}, the confession to it implies a sense of guilt when consuming content on TikTok.

\subsubsection{Policy and Governance}
P25 highlighted the absence of a clear policy in an older niche community that allowed people of all ages to communicate regardless of their age. P9 even humorously recalled how the 1998 chat platform Kilahu~\cite{kilahu2005} served as their first dating platform at the age of 12. These concerns about the policies of older platforms illustrate how the lack of regulation led to fundamental issues in moderation and age restrictions --- problems that future platforms sought to address. The so-called VZ-Networks, as the first nationally acclaimed social networks, created SchülerVZ, a platform specifically for school students, to avoid the moderation challenges faced by older platforms like Kilahu.

With some participants voicing personal discomfort and perceptions of its ``toxicity''~(P6, P16, P25), P1, P3, and P32 mentioned Twitter’s leadership change to Elon Musk and its rebranding to X as reasons for seeking alternative microblogging platforms. In the process migrating away from X, P1 and P3 reported being unable to find a suitable alternative. P32 decided to settle on Bluesky without naming a clear reason for it. 

Although not the focus of the study, international social media realities differ fundamentally from the German participants' perspectives on privacy and security, particularly regarding the policies they must adhere to. Participants who did not engage with the German social media culture shared how government interventions in Iran~\cite{yeeIranInternet2022} blocked the access to 70\% of the internet~(P13, P29).

\section{Discussion: Temporalities of Migration Causes}

Throughout our analysis, we gained an initial understanding of the intricacies involved in digital migration processes between social media platforms. Our findings align with Fiesler and Dym's observations on the complexities of digital migration~\cite{fieslerMoving2020}, particularly within the broader social media context. As people navigate different platforms over time, the alignment between a person's life stage and a platform's features can foster meaningful experiences. For instance, just as Tumblr's sociality and context resonate deeply with the realities of trans individuals~\cite{haimsonTransitionMachinery2018}, the group feature of SchülerVZ – potentially confusing by today's standards – was well-suited to the lives of school students in the pre-Facebook era.

We also echo deVito et al.'s finding that perceptions of the social media landscape are deeply personal and ever-changing~\cite{deVitoTooGay2018}. However, such perceptions are not static; they are shaped by evolving concepts of popularity, security, and the social context in which platforms are experienced. 

\subsection{Peer-Driven Popularity of SchülerVZ and Facebook}

A common pattern among all participants is the migration from SchülerVZ to Facebook, highlighting popularity as a significant factor in choosing a social media platform. However, participants distinguished between two types of popularity: a platform's general, global popularity and its popularity within their community, friends, and family. This distinction reveals that the perception of popularity during migration processes is often shaped by one's peers.

In addition to choosing platforms to communicate with specific groups of people, early platforms frequently served as extensions of real-world social interactions. When transitioning to Facebook, participants rarely reflected explicitly on what was popular among their peers. Instead, they joined Facebook because their peers were already present on a platform perceived as widely used by the general public. This phenomenon appears to be similar to the Matthew effect~\cite{mertonMatthewEffect1968}, where the presence of others on a platform increases its perceived value, leading to further growth and adoption by additional users. As more people migrated to Facebook, it became an even more attractive choice, reinforcing its dominance in social media.

\subsection{Feature Timing and Nostalgic Affordances}

To our surprise, platform features played a minor role to our participants. However, they appreciated them only when they aligned with the timing and context of their lives. For instance, while Snapchat introduced the popular Stories feature, participants primarily associated it with Instagram. This shift suggests that the adoption and impact of certain features are driven by social rather than technical factors, and depend on user engagement with the platform as a whole.

Contemporary social media features played a comparatively minor role for participants compared to those from now-defunct platforms. We argue that the act of reminiscence prompted participants to reflect positively on the early days of the internet~\cite{alizadehReminscence2022}, valuing features they associate with joyful memories. This highlights how different life stages influence users' perceptions and interactions with newer platforms compared to older ones. The value of a platform feature is shaped not only by its utility but also by the social and personal context in which it is experienced.

\subsection{Local Perceptions of Security}

Digital migrations toward more secure platforms were particularly evident in participants' moves to alternatives to WhatsApp. The misconception of Telegram as a secure messaging applications~\cite{greenTelegram2024} and the participants' consistent pursuit of the most secure communication platform reveals how maleable perceptions on platform factors could potentially be. It opens the question of the factors influencing opinion formation on platform traits from a social perspective. Nevertheless, the consistent motivation to prioritize secure communication has remained a driving factor throughout the history of messaging apps.

\subsection{Limitations and Future Work}

Our study showcased the local and regional factors that influence the reasons for digital migration. Consequently, we must further observe the differences of multiple locations to achieve a broader understanding of the phenomena. 
To synthesize our findings, we are conducting follow-up interviews and a parallel quantitative study using the codes of our analysis as a starting point. 

\section{Conclusions}
In this paper, we contribute the first qualitative insight into the locality and temporality of social media migration. We conducted a study using a novel qualitative data gathering approach to map out the social media journeys our participants have taken. Through our qualitative content analysis, we show how social media features need to align with its audience's sociality and life stage to have a considerable effect on migrating to social media platforms in the future. We hope to provide a valuable reference to researchers on social media realities and the sociotechnical impact of platform design.

\balance
\bibliographystyle{ACM-Reference-Format}
\bibliography{submission}
\end{document}